\documentclass[prl,twocolumn]{revtex4}

\usepackage{epsfig}
\usepackage{latexsym}
\usepackage{amsfonts}
\usepackage{amsmath}
\usepackage{amsthm}
\usepackage{amssymb}
\usepackage{amsbsy}
\usepackage{multirow}

\newsavebox{\uuunit}
\sbox{\uuunit}
    {\setlength{\unitlength}{0.825em}
     \begin{picture}(0.6,0.7)
        \thinlines
        \put(0,0){\line(1,0){0.5}}
        \put(0.15,0){\line(0,1){0.7}}
        \put(0.35,0){\line(0,1){0.8}}
       \multiput(0.3,0.8)(-0.04,-0.02){12}{\rule{0.5pt}{0.5pt}}
     \end {picture}}

\def\be{\begin{equation}}
\def\ee{\end{equation}}
\def\bea{\begin{eqnarray}}
\def\eea{\end{eqnarray}}

\newcommand{\beq}{\begin{eqnarray}}
\newcommand{\eeq}{\end{eqnarray}}

\def\L{\Lambda}

\def\f{\phi}

\def\m{\mu}

\def\p{\pi}
\def\r{\rho}

\def\pa{\partial}

\def\nonu{\nonumber \\{}}
\def\half{{1 \over 2}}
\def\sF{{{ F}\!\!\!\!\hskip.8pt\hbox{\raise1pt\hbox{/}}\,}}
\def\som{{{ \omega}\!\!\!\!\hskip.8pt\hbox{\raise1pt\hbox{/}}\,}}
\def\sJ{{{\rm J}\!\!\!\!\hskip.8pt\hbox{\raise1pt\hbox{/}}\,}}

\begin{document}

\title{Relating chronology protection and unitarity through holography}
\author{Joris Raeymaekers}
\affiliation{Institute of Physics of the ASCR, v.v.i.,
Na Slovance 2, 182 21 Prague 8, Czech Republic}
\author{Dieter Van den Bleeken}
\affiliation{NHETC and Dept. of Physics and Astronomy, Rutgers
University, Piscataway, NJ 08855, USA}
\author{Bert Vercnocke}
\affiliation{Institute for Theoretical Physics, K.U.Leuven, Celestijnenlaan 200D, B-3001 Leuven, Belgium}
\date{\today}
\begin{abstract}
We give a simple nonsupersymmetric example in which chronology
protection follows from unitarity and the AdS/CFT correspondence.
We consider a ball of homogeneous, rotating dust in global AdS$_3$
whose backreaction produces a region of G\"odel space inside the
ball. We solve the Israel matching conditions to find the geometry
outside of the dust ball and compute its quantum numbers in the
dual CFT. When the radius of the dust ball exceeds a certain
critical value, the spacetime will contain closed timelike
curves. Our main observation is that precisely when  this critical
radius is exceeded, a unitarity bound in the dual CFT is violated,
leading to a holographic argument for chronology protection.
\end{abstract}
\maketitle


Kurt G\"odel  was the first to emphasize that Einstein's equation,
in the presence of seemingly innocuous matter sources, can lead to
causality violating geometries containing closed timelike curves
(CTCs) \cite{Godel:1949ga}. Since then, classical solutions with
CTCs have popped up ubiquitously, including supersymmetic versions of G\"odel space
in supergravity theories, both in 3+1 dimensions as well as in
their higher-dimensional parent theories
\cite{Gauntlett:2002nw}. Such
spacetimes lead to a variety of pathologies, both within classical
general relativity as well as for interacting quantum fields
propagating on them (see \cite{Friedman:2008dh} for a review and
further references). This led Hawking to propose the Chronology
Protection Conjecture, stating that regions containing CTCs cannot
be formed in any physical process \cite{Hawking:1991nk}. It is
expected \cite{Kay:1996hj} that a fully consistent treatment of
such a dynamical argument behind chronology protection requires
the issue to be addressed in a quantum theory where both matter
and gravity itself are quantized.

The AdS/CFT correspondence \cite{Maldacena:1997re} proposes that
combined quantum gravity and matter systems on anti-de-Sitter
(AdS) spaces have a holographic dual description in terms of a
unitary conformal field theory (CFT) in one lower dimension. It is
therefore ideally suited to study the issue of chronology
protection in asymptotically AdS spaces. Indeed, several examples
are known \cite{Herdeiro:2000ap} where the appearance of CTCs in a
BPS sector of the bulk theory is quantum mechanically forbidden as
it would correspond to the violation of a unitarity bound in the
dual CFT. In this Letter, we show that a similar conclusion holds
for 2+1 dimensional G\"odel space  and give a simple argument that
creating a patch of G\"odel space large enough  to contain CTCs
would require violating unitarity in the dual CFT. An important
novel feature of our example is that it doesn't rely on
supersymmetry but only on the  general properties of gravity
theories on AdS$_3$ that were established in
\cite{Brown:1986nw,Strominger:1997eq}. In particular, our argument
only relies on the fact that, in a unitary CFT, all states have
nonnegative conformal weights.

Let us now outline our argument. G\"odel's original 3+1
dimensional solution  is the product of a nontrivial 2+1
dimensional space and a line, and we will  here consider only the
2+1 dimensional part, henceforth referred to as G\"odel space.
G\"odel space is a solution of 2+1 dimensional anti-de-Sitter
gravity with a source of homogeneous  rotating dust. The dust
needs to rotate in  AdS$_3$ in order to be stationary, since
otherwise it would collapse to form either a conical defect or a
BTZ black hole depending on its total mass
\cite{Ross:1992ba,Vaz:2008uv}. We will consider a 2 dimensional
ball (i.e. a disc) of such rotating dust placed in global AdS$_3$
and give a detailed analysis of the resulting geometry. The metric
inside the ball is that of G\"odel space, and the one outside a
generalized BTZ metric \cite{Banados:1992wn} describing an object
with mass and angular momentum which we determine by solving a
matching problem (a similar 3+1 dimensional problem was considered
in \cite{Bonnor:1997wz}). The AdS/CFT dictionary then tells us the
quantum numbers of the dust ball in the dual CFT. When we vary the
radius of the dust ball while keeping the energy density of the
dust constant, CTCs appear when the radius exceeds a critical
value.  The main question we want to address is what this critical
radius corresponds to in the dual CFT. We will see that it
corresponds precisely to the unitarity bound stating that
conformal weights in the CFT have to be nonnegative, implying that
the formation of a dust ball containing CTCs is forbidden by
unitarity.

\section{The stationary dust ball solution}

We will consider a combined gravity and matter system in AdS$_3$,
where we will not specify  the microscopic matter content in
detail. We assume that the matter sector 
 can effectively produce a source of pressureless dust. Hence we will consider Einstein's
equation with negative cosmological constant $\L = -1/l^2$: \be
R_{ab} - \half R g_{ab}- \frac{1}{l^2}g_{ab}= 8 \p G T_{ab} \,
,\label{einsteineqns} \ee where $G$ is the 2+1 dimensional Newton
constant, $l$ is the AdS$_3$ radius and we take \be T_{ab}
=\frac{\rho}{2\pi G l^2} u_a u_b\, , \label{dustsource} \ee with
$u$ a unit timelike vector and $\rho$ a dimensionless
number parametrizing the energy density. 

We will now solve (\ref{einsteineqns}) for a homogeneous ball of
rotating dust, where we take the energy density $\rho$ to be
nonzero and constant inside the ball and zero outside. Inside the
ball, the metric will be that of G\"odel space, while outside
 we expect a metric of generalized  BTZ type characterized by a mass $M$ and angular momentum $J$.

This leads to  the following ansatz for our matching problem. On
the inside, we have the G\"odel  space metric \be
ds_-^2=l^2\left[-(d t+ \m\frac{ r^2}{(1- r^2)}d\f)^2 + \m\frac{d
r^2+ r^2d\f^2}{(1- r^2)^2}\right]\,.\label{Godelcoords} \ee where
$r$ runs between 0 and $r_0 \leq 1$, the radius where the dust
region ends. The angular coordinate $\phi$ is identified with
period $2 \pi$. The Einstein equations
(\ref{einsteineqns}),(\ref{dustsource}) determine $\mu$ in terms of
the density of the dust as \be \mu ={ 1\over 1-\r
}\label{enden} \, . \ee The physical values are $\rho \geq 0$, for
positive energy, and $\rho
<1$, for a Minkowski signature of the resulting metric. Note that for $\m = 1$, $(\rho=0)$, the metric describes
global AdS$_3$.  When $r_0$ exceeds the critical value $1/\sqrt{\m}$, CTCs appear since
$\pa_\phi$ becomes a timelike vector in the region $r>
1/\sqrt{\m}$.

Outside of the dust ball, we take take a metric ansatz which is  a
vacuum solution to (\ref{einsteineqns})  and  which generalizes
the BTZ metric: \bea ds_+^2 &=& l^2\left[ -  (u - M) d \tilde t ^2
+ J d \tilde t d \tilde \f + u d \tilde \f^2 + { d u^2 \over 4 f(u)}
\right]\,, \nonu f(u) &=&  u^2 - M u + { J^2 \over 4}. \label{genBTZ}
\eea The angle $\tilde \phi$ is identified with period $2\p$ and
the real parameters $M,\ J$ are the ADM mass and angular momentum
(in convenient units) respectively. The
function $f$ is related to the determinant of the  induced metric
$h^+_{ij}$ on a surface of constant $u$ by $\det h^+ = - f$. Let
us review the properties of this class of  metrics in the various
regions of $(J,M)$ parameter space.

In the  region $M^2 \geq J^2,\ M \geq 0$, which we will call
region I, the metrics (\ref{genBTZ}) describe BTZ black holes
\cite{Banados:1992wn}. The function $f$ has two positive real
zeroes, which correspond to the inner and outer horizons. Unlike
their 3+1 dimensional cousins, BTZ black holes have no curvature
singularities
. Instead, in the region  $u<0$ the
space contains CTCs. This `singularity in the causal structure' is
hidden behind a horizon. In the region $u \geq 0$,  the standard
radial BTZ coordinate is related to $u$ as $u = r^2_{\rm BTZ}$.

For $M^2 \geq J^2,\ M < 0$, henceforth referred to as region II,
the metric describes a spinning conical defect. The function $f$
has two negative real zeroes, between which the signature of the
metric becomes Euclidean. One can verify that at the largest zero
$u_+$, the metric has a conical singularity arising from a
pointlike source. The range of the
$u$-coordinate is $u \geq u_+$ and, as before, there are CTCs in
the region where $u_+ \leq u<0$. Both the CTC region and the
defect singularity are `naked' and not hidden behind a horizon.
 There is one exceptional point,
namely $M = -1,\ J=0$, for which the geometry becomes smooth global AdS$_3$.
This special point corresponds to the conformally invariant vacuum state in the dual CFT.

And finally, for $M^2 < J^2$, denoted by region III, the metric
describes an overspinning object. The function $f$ has no real
zeroes and the metric is free of curvature singularities. The
range of $u$ is the real line and the space contains a `naked' CTC
region for negative values of $u$.

Let us now briefly discuss part of the  AdS/CFT dictionary. The
Virasoro quantum numbers of the spaces (\ref{genBTZ}) can be
extracted following the standard procedure of computing the
renormalized boundary stress tensor and extracting its Fourier
coefficients \cite{Balasubramanian:1999re}. One finds that these
spaces correspond to states with conformal weights \bea
L_0 &=&{c \over 24} ( M+ J + 1 )\,,\\
\bar L_0  &=&{c \over 24} ( M-J +1)\,. \eea The central charge of the
CFT is given by \cite{Brown:1986nw} \be c = { 3 l \over 2 G}\,.
\label{BH} \ee 
Unitarity implies that conformal weights in the CFT are positive,
leading to the bound $L_0 \geq 0, \bar L_0 \geq 0$. In terms of
$M$ and $J$, this is equivalent to \be M+1 \geq
|J|\,.\label{unitbound} \ee States violating this bound are forbidden
by unitarity and, according to the AdS/CFT conjecture, cannot be
part of the spectrum in a consistent quantum gravity theory on
$AdS_3$.

We will now match the inside metric for $ r \leq r_0$ to the
outside  metric for $ u\geq u_0$ for arbitrary values of the
parameters $\r$ and $r_0$. Since both metrics have a single $2 \p$
identification on the coordinates $\f, \tilde \f$ respectively,
the coordinates $t,\f$ and $\tilde t , \tilde \f$ have to be
related as follows \bea
t &=& c_1 \tilde t\,, \label{time}\\
\f &=& c_2 \tilde t + \tilde \f \label{ang}, \eea with $c_1,c_2$
two constants.

The Israel matching conditions \cite{Israel:1966rt} require that
the metric and the extrinsic curvature are continuous across the
edge of the dust ball \footnote{The singularity in higher curvature corrections due to
the sharp edge of our dust ball can be resolved by smoothening the density over a length scale $L$. If $L$ is large
compared to $G$ but small compared to $l$,  both curvature
corrections (suppressed by $G/L$) and deviations from our
solution (suppressed by $L/l$) remain small.}: \bea
h^-_{ij} &=& h^+_{ij}\,,\label{cont}\\
K^-_{ij} &=& K^+_{ij}\,.\label{matchingcond} \eea Here $h^-_{ij}$
($h^+_{ij}$) is the induced metric on the $r = r_0$ $(u=u_0)$
boundary surface and $K^\pm_{ij}$ are the corresponding extrinsic curvatures.
The conditions (\ref{cont}),(\ref{matchingcond}) give 6 equations
for the five undetermined parameters $M,J,u_0, c_1,c_2$ in terms
of the two physical parameters $\rho, r_0$ of the ball of dust.
After some algebra one can check that there are 2 solutions: \bea
J &=& \pm \frac{2\,\rho\, r_0^4 }{(1-r_0^2)^2 (1-\rho )^2}\label{J}\,,\\
M &=& -\frac{(1-\rho )^2-2 r_0^2 \left(1-\rho ^2\right)+r_0^4 \left(1+\rho ^2\right)}{(1-r_0^2)^2 (1-\rho )^2}\label{M}\,,\\
u_0 &=&  \frac{r_0^2 (1-\r-r_0^2)}{(1-r_0^2)^2 (1-\rho )^2}\label{u0}\,,\\
c_1 &=&
\pm \left(1-\frac{2 r_0^2 \r}{\left(1-r_0^2\right) (1-\rho )}\right)\label{c1}\,,\\
c_2&=&\pm 1\,. \label{c2} \eea The two solutions have opposite
angular momentum and are related by a change of sign for the
angular coordinates $\phi,\ \tilde \phi$. We will fix this freedom
in what follows by taking $J$ to be positive, choosing the
positive sign in these equations.

\section{Discussion}

Let us now discuss the physical properties of the matched
solution (\ref{J})-(\ref{c2}). First we examine how the $(r_0,
\r)$ parameter space is mapped into the $(J,M)$ plane. As
mentioned before $(r_0,\r)$ take  values in $[0,1)\times[0,1)$.
Their relation to $(J,M)$ is clearly not onto and also not
injective, as the Jacobian of the map vanishes at
$\r=\frac{1-r_0^2}{1+r_0^2}$. It turns out that there are two
branches in the $(J,M)$ plane, depending on whether the dust
density $\rho$ is greater or smaller than
$\frac{1-r_0^2}{1+r_0^2}$.
\begin{figure}[h!]
\centering{
\includegraphics[width=150pt]{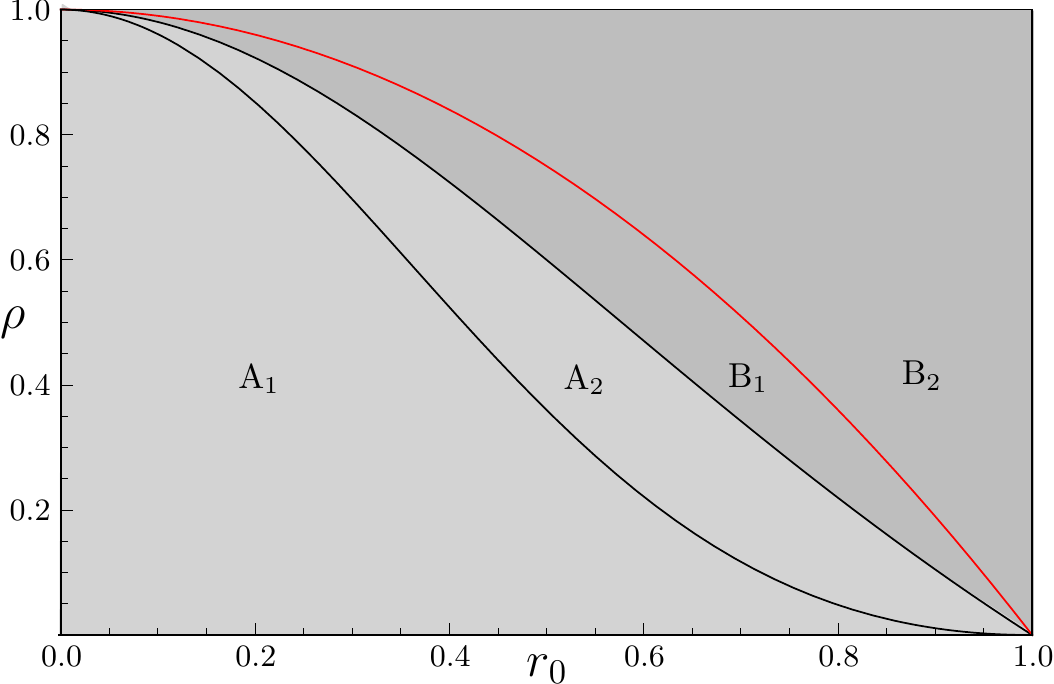}
}
\vspace{.5cm}
\centering{
\includegraphics[width=150pt]{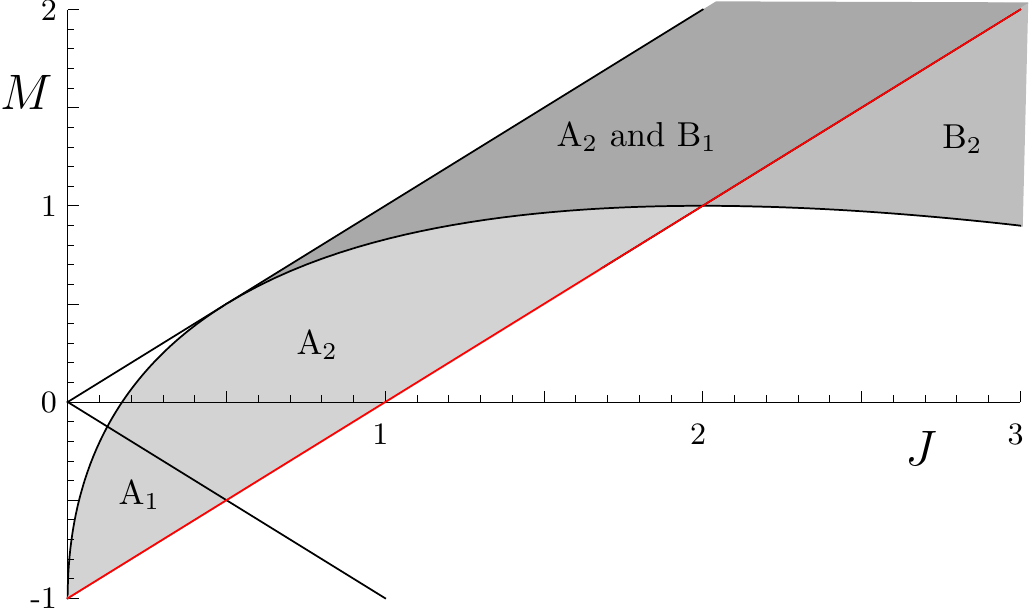}
} \caption{The relation between the physical parameter space
$(r_0, \rho)$  of the dust ball (upper figure) and the values
$(J,M)$ of the outside metric (lower figure). Observe that the region B$_2$, where the solutions contain closed timelike curves lies in the half-plane $J>M+1$, where unitarity is violated. \label{jmfig}}
\end{figure}
More precisely the two branches are
\begin{itemize}
 \item Branch A $\left(\r \leq\frac{1-r_0^2}{1+r_0^2}\right)$:
\be \begin{cases}  J-1\leq M\leq-1+2\sqrt{2J}-J&\mbox{when}\quad
J\leq\frac{1}{2}\,,\\ J-1\leq M\leq J &\mbox{when}\quad
J\geq\frac{1}{2}\,.\end{cases} \ee \item Branch B
$\left(\r\geq\frac{1-r_0^2}{1+r_0^2}\right)$: \be
-1+2\sqrt{2J}-J\leq M\leq J\,. \ee
\end{itemize}
These branches are shown  in figure \ref{jmfig} and we will now
discuss them in more detail.

Within branch A, for values of $\r \leq\left(\frac{1-r_0^2}{1+r_0^2}\right)^2$ (i.e.  in
the region A$_1$ in figure \ref{jmfig}), the outside metric is of
type II (conical defect). Note that there are two limits where the
outside becomes the AdS$_3$ vacuum. One is simply setting $r_0=0$.
There is only an outside space in this case, as $u_0=0$ and
$M=-1,\ J=0$. This outside space covers all of global AdS$_3$. The
second limit is taking the energy density of the dust to be zero,
$\rho=0$. Now both the inside and outside metric become a patch of
global AdS$_3$, albeit in different coordinate systems. The gluing
conditions (\ref{ang}), (\ref{u0}) simply reduce to the
appropriate coordinate transformation relating the two coordinate
systems. For  values of $\r$ and $r_0$ lying in the region
A$_2$ in figure \ref{jmfig}, the outside metric is of type III
(overspinning  object). Within branch B, the outside metric is
always also of type III. On the line $\r = \frac{1-r_0^2}{1+r_0^2}$ where branch A meets
branch B, $M$ becomes equal to $J$ and the outside metric is
formally of type I, describing an extremal BTZ black hole. The
edge of the dust ball $u_0$ coincides precisely with the black hole horizon. Since on this line $c_1$ becomes zero
(\ref{c1}), the redshift between the AdS time and the G\"odel time becomes infinite and the glueing singular. Already in \cite{Vaz:2008uv} it was observed that glueings to a dust region cannot give rise to outside metrics of type I.

Now we address the issue of closed timelike curves in our
matched solutions. Remember that a priori we have two regions  in
which closed timelike curves can appear in the solution. The
inside part of the metric (G\"odel space) has CTCs when
$r_0^2>\frac{1}{\m}=1-\rho$, the outside metric when $u_0<0$.
But observe that by (\ref{u0}) these two conditions are
equivalent, hence either no closed timelike curves appear at all,
or they appear both in the inside and outside parts of the metric.
In parameter space these closed timelike curves can only appear in part of
branch B, wich is denoted as region
B$_2$ in figure \ref{jmfig}. Hence on branch B, CTCs can be
made to appear by smoothly varying the parameters $(\r ,r_0)$,
while the solutions remain seemingly well behaved in all other
respects.

Now let us discuss the unitarity bound (\ref{unitbound}), which
gives an extra constraint on which outside metrics are physically
acceptable. On branch A, all the outside metrics satisfy the
unitarity bound, while branch B is divided into a region where the
bound is satisfied and one where it is violated. In fact, the
bound for the absence of CTCs $r_0^2 \leq 1-\rho$  precisely
coincides with the unitarity bound $M+1\geq |J|$. This can be seen
directly as by (\ref{J}), (\ref{M}) \be M+1-|J|=\frac{4
\rho\,r_0^2 (1-\rho -r_0^2 )}{(1-r_0^2)^2 (1-\rho )^2}\,.\ee Hence
the condition of unitarity is equivalent to that of the absence of
closed timelike curves. 


\section{Outlook}
In this Letter, we have discussed an example where the appearance
of CTCs in a G\"odel region within  AdS$_3$ was shown to precisely
coincide with the violation of a unitarity bound in the dual CFT.
Based on our result and other examples in the literature
\cite{Herdeiro:2000ap}, it would be natural to propose an AdS
version of the Chronology Protection Conjecture, stating that
regions with CTCs in AdS spaces cannot be formed as a result of
any unitary process. The AdS/CFT correspondence could  in
principle be used to address whether this proposal is true in
general. If so, it would be very interesting to gain insight into
the deeper dynamical mechanism that prevents the formation of
regions with CTCs, see e.g. \cite{Drukker:2003sc} for some
proposals in the context of string theory.

\section{Acknowledgments}
\acknowledgements{ We would like to thank  M. Caldarelli and M.
Headrick for valuable discussions.

This work was supported by the EURYI grant EYI/07/E010 from
EUROHORC and ESF (JR), by the DOE under grant DE-FG02-96ER40949
(DVdB), by the Federal Office for Scientific,  Technical and
Cultural Affairs through the Interuniversity Attraction Poles
Programme Belgian Science Policy P6/11-P and by the project
G.0235.05 of the FWO-Vlaanderen (BV).  BV is an Aspirant of the
FWO-Vlaanderen.}

\end{document}